# Dark states and delocalization: competing effects of quantum coherence on the efficiency of light harvesting systems

Zixuan Hu[1,2], Gregory S. Engel[3], Fahhad H. Alharbi[2] and Sabre Kais*[1,2]

1. Department of Chemistry, Department of Physics, and Birck Nanotechnology Center, Purdue University, West Lafayette, IN 47907, United States
2. Qatar Environment and Energy Research Institute, College of Science and Engineering, HBKU, Doha, Qatar
3. Department of Chemistry, James Franck Institute and the Institute for Biophysical Dynamics, University of Chicago, Chicago, IL 60637, United States

*Email: kais@purdue.edu

Natural light harvesting systems exploit electronic coupling of identical chromophores to generate efficient and robust excitation transfer and conversion. Dark states created by strong coupling between chromophores in the antenna structure can significantly reduce radiative recombination and enhance energy conversion efficiency. Increasing the number of the chromophores increases the number of dark states and the associated enhanced energy conversion efficiency, yet also delocalizes excitations away from the trapping center and reduces the energy conversion rate. Therefore, a competition between dark state protection and delocalization must be considered when designing the optimal size of a light harvesting system. In this study, we explore the two competing mechanisms in a chain-structured antenna and show that dark state protection is the dominant mechanism, with an intriguing dependence on the parity of the number of chromophores. This dependence is linked to the exciton distribution among eigenstates which is strongly affected by the coupling strength between chromophores and the temperature. Combining these findings, we propose that increasing the coupling strength between the chromophores can significantly increase the power output of the light harvesting system.

## I.     Introduction

Natural and artificial light harvesting systems exploit electronic coupling of identical chromophores to generate efficient and robust excitation transfer and conversion. Recent probes of quantum coherence in light harvesting systems have enabled us to study not only the couplings, but also how the environment surrounding the chromophores drives photosynthetic energy transfer and conversion[1-8]. For many photovoltaic devices, radiative recombination arising from the principle of detailed balance limits efficiency. Indeed, radiative recombination is one of the fundamental contributors to the famous Shockley-Queisser limit[9]. Previous works by Scully and coworkers[10, 11] have shown that quantum coherence induced by either microwave or noise can break the detailed balance and reduce the radiative recombination rate. In artificial models inspired by natural light harvesting complexes, quantum states formed by coupling between multiple chromophore subunits in the antenna system have been shown to prevent radiative recombination through dark state protection and effectively increase the energy transfer efficiency to the trapping center[12-14]. On the other hand, the formation of the collective eigenstates delocalizes the excitation away from the trapping site, and may reduce the energy transfer efficiency. Therefore, a competition between dark state protection and delocalization needs to be considered when optimizing the size of light harvesting systems[15]. Here, we examine these two competing mechanisms in detail for an antenna composed of a chain of varied numbers of chromophore subunits. We show that dark state protection is the dominating mechanism in the presence of intraband transitions mediated by phononic dissipation. Surprisingly, we observe an intriguing dependence on the parity of the number of subunits. The resultant plot of energy transfer power over the number of dipoles shows a zigzag pattern in which the energy transfer efficiency depends on whether the number of dipoles is even or odd. In the absence of intraband transitions, dark state protection is turned off and the delocalization effect becomes visible. Combining these findings, we propose a way to increase the energy transfer efficiency of the light harvesting system by increasing the coupling strength between the dipoles and using an even number of dipoles on the antenna such that dark state protection is maximized.

## II. Model

Figure 1 shows the model used in this study. The model includes an antenna system that receives light energy to produce optical excitations and an energy trapping site that converts excitations transferred from the antenna into power.

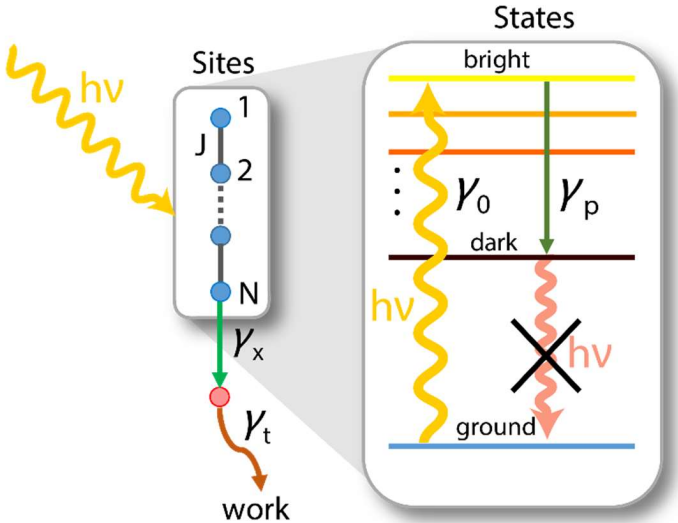

*Figure 1. Model showing (left) the physical arrangement of the chromophores (structures in the box) with the trapping site (red dot), and (right) the energy diagram for the antenna. The bright state is excited by light to create an exciton, which then moves down to the dark state through phononic dissipation. The dark state prevents radiative recombination, thereby improving exciton transfer efficiency to the trapping site.*

The antenna system consists of a chain of $N$ identical two-level optical emitters coupled through nearest-neighbor dipole-dipole interactions, whose Hamiltonian is given by:

$$H_a = \omega \sum_{i=1}^{N} \sigma_i^+ \sigma_i^- + J \sum_{i=1}^{N-1} (\sigma_i^+ \sigma_{i+1}^- + \sigma_{i+1}^+ \sigma_i^-) \tag{1}$$

where $\hbar = 1$, $\omega$ is the site energy, and $J$ is the coupling strength. The use of the Pauli raising and lowering operators explicitly ensures that a single site cannot support more than one excitation. The single-excitation eigenstates can be solved easily:

$$|\psi\rangle_k = \sqrt{\frac{2}{N+1}} \cdot \sum_{j=1}^{N} \sin\left(\frac{k\pi}{N+1} j\right) c_j^\dagger |0\rangle \tag{2}$$

where $k$ is an integer ranging from 1 to $N$. The eigenenergies of $|\psi\rangle_k$'s are given by:

$$E_k = \omega + 2J \cos\left(\frac{k\pi}{N+1}\right) \tag{3}$$

The antenna system is connected to the trapping site through the $N^{th}$ dipole. To calculate the steady state dynamics of the combined antenna-trap system, we follow the procedures used in Ref.[14, 15] by setting up a standard Lindblad optical master equation:

$$\dot{\rho} = -i[H_a + H_t, \rho] + D_o[\rho] + D_p[\rho] + D_t[\rho] + D_x[\rho] \tag{4}$$

where $\rho = \rho_a \otimes \rho_t$ is the total density operator of the antenna-trap system, $H_t = \omega_t \sigma_t^+ \sigma_t^-$ is the trapping site Hamiltonian, $D_o[\rho]$ is the optical dissipator describing the interband transitions between different excitation levels, $D_p[\rho]$ is the phononic dissipator describing the intraband transitions within one excitation level, $D_t[\rho]$ describes the decay process of the trapping site, and $D_x[\rho]$ describes the extraction process from the antenna ring to the trapping site. The parameters associated with each of the dissipators in equation (4) are given in Table 1 with the physical meaning of each parameter described in the leftmost panel. For the detailed equation forms where these parameters appear in equation (4) please see the appendix.

| Parameter | Symbol | Value |
| --- | --- | --- |
| Antenna site energy | $\omega$ | 1.76 eV |
| Antenna coupling strength | $J$ | 20 meV |
| Antenna optical decay rate | $\gamma_o$ | 0.001 meV |
| Antenna phononic decay rate | $\gamma_p$ | 1 meV |
| Trap optical decay rate | $\gamma_t$ | $10^{-4}$ meV |
| Antenna to trap extraction rate | $\gamma_x$ | $10^{-5}$ meV |
| Ambient temperature | $T$ | 300 K |
| Optical temperature | $T_o$ | 5800 K |

*Table 1. Parameters used in the numerical calculations. For detailed equation forms where these parameters appear in equation (4) please see the appendix.*

Steady state solution of $\rho = \rho_a \otimes \rho_t$ in equation (4) is obtained using the open-source quantum dynamics software QuTiP[16]. The theory of quantum heat engine[11, 17] is then used to calculate the power output of our light harvesting system, for which the current is given by $I = e\gamma_t \langle \rho_{te} \rangle_{ss}$, and the voltage is given by $eV = \hbar\omega_t + k_B T \ln\left(\frac{\langle \rho_{te} \rangle_{ss}}{\langle \rho_{tg} \rangle_{ss}}\right)$, with $\langle \rho_{te} \rangle_{ss}$ being the steady state population of the trap's excited state, $\langle \rho_{tg} \rangle_{ss}$ the steady state population of the trap's ground state, $e$ the fundamental charge, $k_B$ the Boltzmann constant, and $T$ the thermal temperature.

## III. Results and discussion

With both the analytic and numerical models established, we are now ready to calculate the power output $P = IV$ for various numbers of dipoles on the antenna ($N$ = 1 to 7). The results are shown in Figure 2.

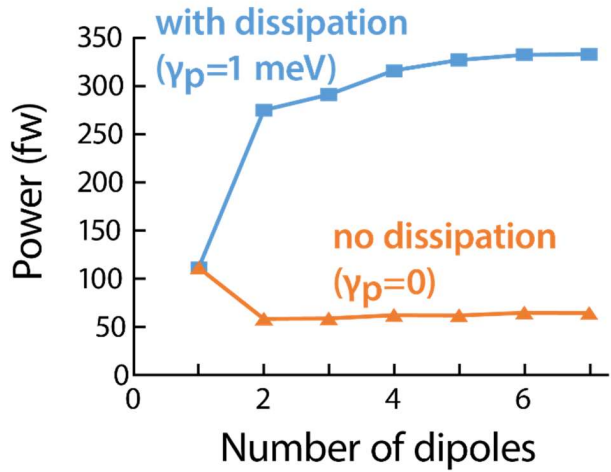

Figure 2. Exciton transfer power calculated with J = 20 meV and T = 300 K with and without dissipation. Other parameters are listed in Table 1. In the presence of dissipation, energy transfer power increases with increasing numbers of coupled dipoles. The most pronounced increase is seen between a single emitter and two emitters. In the absence of dissipation, a decrease in power is observed between a single emitter and two emitters. Further increases in the number of dipoles do not lead to an obvious trend.

In Figure 2, it is clear that the exciton transfer dynamics depend critically on phononic dissipation. When dissipation is present, the power of exciton transfer increases many-fold. When no dissipation is present, power drops by half from a single dipole to a double dipole chain and remains at this low level as the number of emitters increases. In both cases, delocalization is present. This delocalization causes the power to drop when dissipation is absent. When dissipation is present, the dark states become populated by dissipation from the bright states and radiative recombination is reduced. This dark state mechanism dominates over the delocalization effect. Below, we focus on the case with dissipation and discuss the case without dissipation in the appendix. Interestingly, if we increase the coupling between dipoles from $J=10$ meV to $J=100$ meV, a zigzag pattern emerges, as shown in Figure 3:

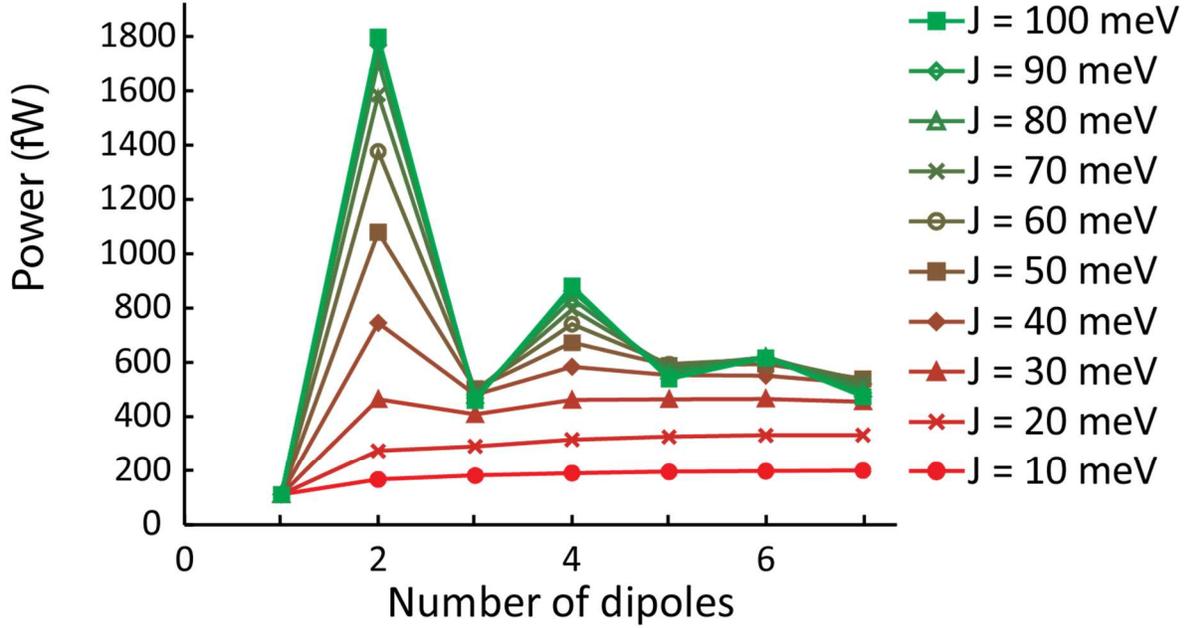

*Figure 3. Exciton transfer power calculated at T = 300 K as coupling between dipoles increases from J = 10 meV to J = 100 meV in 10 meV increments. Other parameters are listed in Table 1. As the coupling strength increases, the efficiency of energy transfer rises while the zigzag pattern becomes more pronounced. Even numbers of dipoles show improved efficiency as compared to odd numbers.*

In Figure 3, we observe that as the coupling strength $J$ increases, the power trend displays a conspicuous zigzag pattern as the number of dipoles ($N$) increases. In addition, for the even $N$'s, there is a considerable power enhancement when $J$ increases; for the odd $N$'s, the power increases with $J$ initially but quickly converges beyond $J=40$ meV. To explain these results, we examine the optical coupling of a single-excitation eigenstate $|\psi\rangle_k$ to the ground state $|0\rangle$:

$$\Gamma_{k0} = \left|\langle\psi_k|\sum_{j=1}^{N}\sigma_j^+|0\rangle\right|^2 = \frac{1}{2(N+1)} \cdot \left|\cot\left(\frac{k\pi}{2(N+1)}\right) \cdot \left(1-(-1)^k\right)\right|^2 \quad (5)$$

In our model, $J > 0$ and by equation (3), the lowest energy eigenstate has $k = N$ whose coupling to the ground is:

$$\Gamma_{N0} = \frac{1}{2(N+1)} \cdot \left|\cot\left(\frac{N\pi}{2(N+1)}\right) \cdot \left(1-(-1)^N\right)\right|^2 \quad (6)$$

The population distribution over the single-excitation eigenstates are governed by the Bose-Einstein distribution $n(E_k) = \left(e^{E_k/k_B T} - 1\right)^{-1}$ where the lowest energy state has the highest population. Consequently, if the lowest energy state is completely dark, the dark state enhancement would be greater than that if the lowest energy state is partially dark because in the latter case, there is a small but finite rate for the excitation to undergo radiative recombination. By equation (6), we see that if $N$ is even, $\Gamma_{N0} = 0$ and the lowest energy eigenstate is completely dark; yet if $N$ is odd, $\Gamma_{N0} = \frac{2}{(N+1)} \cdot \left|\cot\left(\frac{N\pi}{2(N+1)}\right)\right|^2 \neq 0$ and the lowest energy eigenstate has finite rate for radiative recombination. With these results, we would expect the energy transfer power of our model light harvesting system to drop for odd $N$'s and rise for even $N$'s, creating a zigzag pattern, exactly as seen in Figure 3. In Figure 3, the zigzag pattern is more noticeable with higher $J$ values because greater coupling strengths between the dipoles increase the energy spacing between different eigenstates, causing more of the population to shift to the lowest energy state. Indeed, if we keep $J = 20$ meV. but vary the temperature from $T = 1$ K to $T = 401$ K, similar energy transfer patterns are obtained (Figure 4).

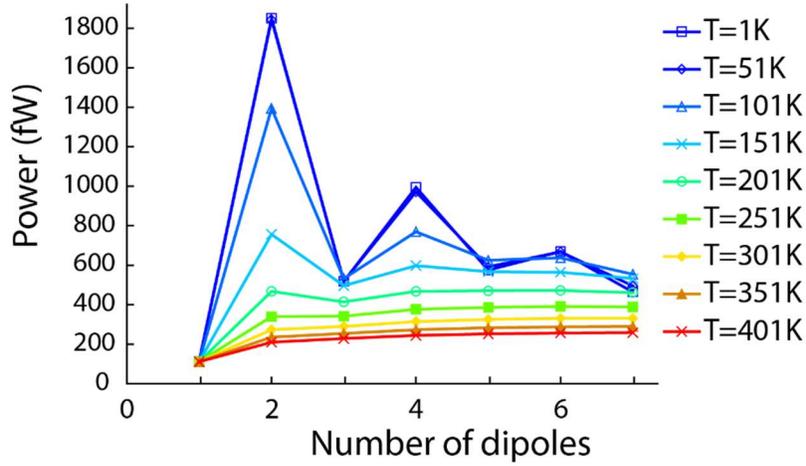

*Figure 4. Exciton transfer power calculated using J = 20 meV and T = 1 K to 401 K, with 50 K increments. Other parameters are listed in Table 1. As the temperature increases, the power transfer efficiency falls. The zigzag pattern is most pronounced at low temperatures, but it persists to ambient temperatures.*

In Figure 4, it is clear that lowering the temperature has a similar effect to increasing the coupling strength, i.e. it pushes the population distribution to the lowest energy state, and the result is an obvious zigzag pattern for temperatures below 151 K. In contrast, when the temperature is increased beyond 200 K, the zigzag pattern becomes less obvious and eventually disappears because, at higher temperatures, the population is evenly distributed across all of the intraband levels. At temperatures above 301 K, the power transfer efficiency at each $N$ decreases monotonically with further increase in temperature, and the unusual alternating pattern almost completely disappears. From this data and the data in Figure 3, we conclude that the effectiveness of the dark state enhancement mechanism depends critically on the population statistics among all the intraband levels. Pushing the population distribution to lower energy levels – either by increasing the coupling strength or by decreasing the temperature – can significantly increase dark state protection and the power enhancement.

Examining the traces with significant zigzag patterns in Figures 3 and 4, i.e. traces with $J$ greater than 0.04 eV or $T$ lower than 151 K, we see that the power efficiency trend for the even $N$'s is different from that for the odd $N$'s. The power for the even $N$'s decreases as $N$ increases, which is due to the lowest energy eigenstate delocalizing away from the trapping site. On the other hand, the odd $N$'s show a power increase as $N$ increases (with the exception of $N = 7$), which is due to the coupling to the ground state $\Gamma_{N0} = \frac{2}{(N+1)} \cdot \left| \cot\left( \frac{N\pi}{2(N+1)} \right) \right|^2$ decreasing over N. The exception of $N = 7$, and for greater $N$'s (not shown), is due to complications caused by the double excitation eigenstates of equation (1). The double excitation states are generated by combining two single excitation states and, hence, the number of double excitation states depends on $N$ as $N \cdot (N-1)/2$. For $N \geq 7$, the number of double excitation states is sufficient to affect the overall dynamics and the power trends differ from what we have discussed so far. This double excitation effect is interesting in its own right and will be examined in a future study. Looking at the separate trends for the odd and even numbers of dipoles, we see again that the effect of the dark state protection mechanism is much more significant than the competing delocalization mechanism: when the effectiveness of the dark state protection differs among the different $N$'s being considered, the trend is determined by the dark state mechanism (the odd $N$ cases); when there is no difference in the effectiveness of the dark state protection among the different $N$'s, the trend is determined by the delocalization mechanism (the even $N$ cases). Going back to Figure 3, an important implication from the results is that by simply increasing the coupling $J$ in the chain structure under room temperature, the power for each $N$ can be greatly enhanced. This inspires a design strategy for maximizing the power for artificial light harvesting systems. Once again, a greater coupling $J$ increases the energy spacing between the single-excitation eigenstates, which pushes the population distribution to lower energy states. The coupling to the ground

$\Gamma_{k0} = \dfrac{1}{2(N+1)} \cdot \left| \cot\left( \dfrac{k\pi}{2(N+1)} \right) \cdot \left(1 - (-1)^k\right) \right|^2$ is greatest when $k = 1$ and with equation (3), the $|\psi_{k=1}\rangle$ state is also the highest in energy. Consequently, pushing the population distribution towards lower eigenstates and away from the brightest state can greatly enhance the output power of the light harvesting system by reducing the exciton recombination rate.

### IV.    Conclusions

In this study, we have examined two competing mechanisms – dark state protection and delocalization – which affect the energy transfer power in a model light harvesting system composed of a chain-structured antenna and a trapping site. Numerical calculations have shown that the dark state mechanism is the dominant mechanism in the presence of intraband transitions. This mechanism creates a curious zigzag pattern on the power versus $N$ plot, amplified by either increasing the coupling between chromophores or lowering the temperature. We attribute the zigzag pattern to the population distribution over the eigenstates, whose optical couplings to the ground state are sensitive to the parity of $N$. We acknowledge the fact that it may be difficult to experimentally observe the zigzag pattern due to the difficultly in controlling the exact length of the chain in the aggregate. Nonetheless, the results presented and the theory proposed clearly illustrate the detailed interplay between the dark state protection and delocalization mechanisms. The results and the proposed theory also provide ideas for designing better artificial light harvesting systems by controlling the exciton population statistics. In terms of future directions, the detailed mechanism of how double excitation eigenstates affect the dynamics and power output for $N \geq 7$ will be studied.

## V. Acknowledgements

This work is supported by the QNRF exceptional grant: NPRPX-107-1-027

## VI. Appendix

### 1. Numerical model

In equation (4) in the main text, there are four dissipators -- $D_o[\rho]$, $D_p[\rho]$, $D_t[\rho]$, and $D_x[\rho]$ -- describing four different physical processes.

$D_o[\rho] = \gamma_o \sum_{\omega_o} \Gamma_{K,K'} \left( N(\omega_o+1) \mathcal{D}[\hat{L}_o, \rho] + N(\omega_o) \mathcal{D}[\hat{L}_o^\dagger, \rho] \right)$ is the optical dissipator describing the interband transitions between different excitation levels, where $\gamma_o$ gives the optical transition rate for the antenna, $\Gamma_{K,K'} = \left| \langle \psi_K | \sum_{j=1}^N \sigma_j^+ | \psi_{K'} \rangle \right|^2$ is the optical coupling strength between two eigenstates of the antenna Hamiltonian with $\omega_o = \varepsilon_K - \varepsilon_{K'} > 0$, $N(\omega_o) = \left( e^{\omega_o/k_B T_o} - 1 \right)^{-1}$ is the optical distribution, and $\mathcal{D}[\hat{L}_o, \rho] = \hat{L}_o \rho \hat{L}_o^\dagger - \frac{1}{2} \{\hat{L}_o^\dagger \hat{L}_o, \rho\}$ is the Lindblad dissipator with $\hat{L}_o^\dagger = |K\rangle\langle K'|$.

$D_p[\rho] = \gamma_p \sum_{\omega_p} \left( N(\omega_p+1) \mathcal{D}[\hat{L}_p, \rho] + N(\omega_p) \mathcal{D}[\hat{L}_p^\dagger, \rho] \right)$ is the phononic dissipator describing the intraband transitions within one excitation level, where $\gamma_p$ gives the phononic relaxation rate, $N(\omega_p) = \left( e^{\omega_p/k_B T_p} - 1 \right)^{-1}$ is the thermal distribution, and $\mathcal{D}[\hat{L}_p, \rho] = \hat{L}_p \rho \hat{L}_p^\dagger - \frac{1}{2} \{\hat{L}_p^\dagger \hat{L}_p, \rho\}$ where $\hat{L}_p^\dagger = |\mu\rangle\langle\nu|$ are the intraband transitions with $\omega_p = \varepsilon_\mu - \varepsilon_\nu > 0$.

$D_t[\rho] = \gamma_t \mathcal{D}[\sigma_t^-, \rho]$ describes the decay process of the trapping site.

$D_x[\rho] = \gamma_x \mathcal{D}[\sigma_N^- \sigma_t^+, \rho]$ describes the extraction process from the antenna ring to the trapping site.

The parameters associated with each of the processes are given in Table 1 in the main text.

### 2. Further discussion on the delocalization effect without dissipation

In the main text, we have studied the dynamics with dissipation where the population distribution over different eigenstates is important. In the following we consider the case without dissipation by setting $\gamma_p = 0$ and focusing on the effect of delocalization. In the orange/triangle curve in Figure 2 in the main text, we observed the power drop from a single dipole to a double dipole chain. Surprisingly, when the number of dipoles is further increased, the delocalization effect does not lead to a monotonic decrease of energy transfer power, but a minor zigzag pattern that cannot be accounted for by the thermal distribution over eigenstates of different degrees of darkness. We show in the following that this behavior can be attributed to the edge state effect caused by the boundary conditions of the chain. Since there is no intraband transition, the only states excited will be the ones accessible from optical transitions, i.e. only the states with finite coupling to the ground state will be populated. The coupling strength to the ground state is calculated in equation (5) in the main text, and in the states with finite coupling, $k$ is odd. For each N in our model, only the terminal site (the $N^{th}$ site) is connected to the trapping site and the probability of finding the excitation on the terminal site for the single-excitation eigenstate labeled by $k$ is given by:

$$p_N(k) = |C_N(k)|^2 = \frac{2}{(N+1)} \cdot \left|\sin\left(\frac{Nk\pi}{N+1}\right)\right|^2 \qquad \text{A(1)}$$

which is plotted in Figure A1 for different $k$ values and dipole number $N$'s:

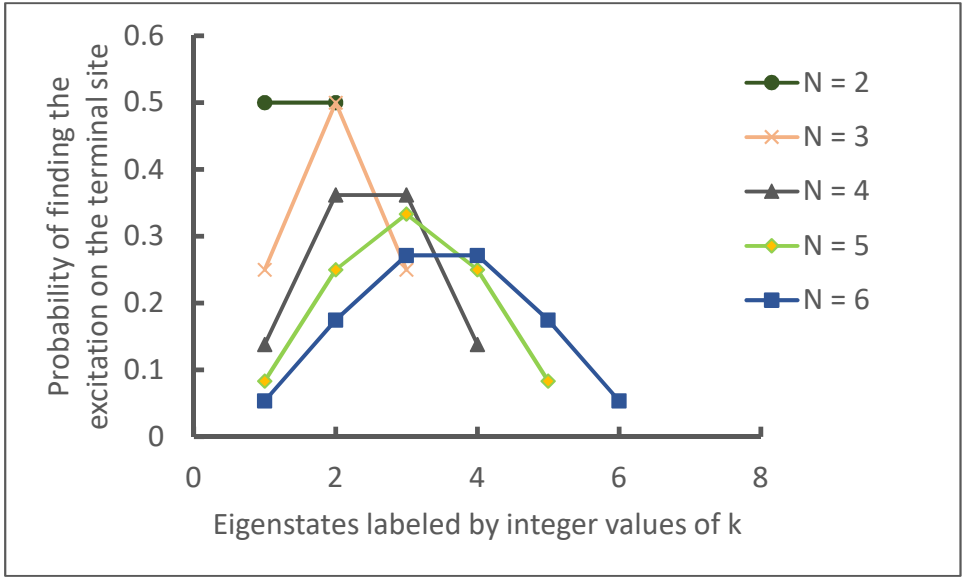

*Figure A1. Probability of finding the excitation on the terminal site for different eigenstates indexed by k. For N greater than 3, the state(s) with the greatest probability of finding the excitation on the terminal site (nearest to the trapping site) lies in the middle of the energy ladder (k values in the middle).*

Figure A1 plots the probability of finding the excitation on the terminal site nearest the trap against different integer $k$ values, which represent the eigenstates of descending energies with increasing $k$. Note that for each $N \geq 3$, there is always a state(s) in the middle of the energy ladder that will have the greatest probability of finding the excitation on the site nearest the trap; this is known as the edge state effect in solid state physics. For example when N = 3, the edge state has $k = 2$ which, by equation (5) in the main text, will not couple to the ground state. Hence, we need to consider the state with the second greatest probability on the $N^{th}$ site, which we call the lesser edge state. For even N's, one of the two edge states will have an odd k, such that it will couple to the ground. Therefore, if we go from e.g. N = 3 to N = 4, the populated eigenstate is actually closer to the trapping site rather than further away from it. In other words, the delocalization effect works in a complicated way due to the existence of edge states.